\documentclass[preprint, 3p, sort&compress]{elsarticle}
\usepackage{dcolumn}
\usepackage{booktabs,tabularx} 
\usepackage{booktabs}
\usepackage{textcomp}
\usepackage{amssymb}
\usepackage{amsthm}
\usepackage[mathlines]{lineno}
\usepackage{graphicx}
\usepackage{units}
\usepackage{url}
\usepackage{amsmath}
\usepackage{amsfonts}
\usepackage{bm}
\usepackage{textcomp}
\usepackage{subfigure}
\usepackage{multicol}
\usepackage{multirow}
\usepackage{verbatim}
\usepackage{rotating}
\usepackage[colorlinks,linkcolor=black,citecolor=red]{hyperref}
\usepackage{float}
\usepackage{epsfig}
\usepackage{dcolumn}
\usepackage{bm}
\usepackage{color}
\usepackage{pstricks}
\usepackage{pst-node}
\usepackage{times}
\usepackage{indentfirst}
\usepackage[english]{babel}
\usepackage{epstopdf}
\addto{\captionsenglish}{%

}
\journal{Physics Letters B}


\usepackage[font=footnotesize]{caption}

\usepackage{overpic}

\newcommand{\mev}{\,\mbox{MeV}}
\newcommand{\mevc}{\,\mbox{MeV}/c}
\newcommand{\mevcc}{\,\mbox{MeV}/c^2}

\newcommand{\gev}{\,\mbox{GeV}\,}
\newcommand{\gevc}{\,\mbox{GeV}/c}

\newcommand{\jpsi}{J/\psi}
\newcommand{\pip}{\pi^+}
\newcommand{\pin}{\pi^-}

\newcommand{\lam}{\Lambda}
\newcommand{\lamb}{\bar{\Lambda}}
\newcommand{\lambb}{\Lambda\bar{\Lambda}}

\newcommand{\sigg} {\Sigma^{0}\bar{\Sigma}^{0}}
\newcommand{\sig} {\Sigma^{0}}
\newcommand{\sigb}{\bar{\Sigma}^{0}}

\newcommand{\ee}{e^+e^-}

\biboptions{numbers,sort&compress}
\begin{document}
\begin{frontmatter}
\title{\boldmath \bf Measurement of the $\ee\to\Sigma^{0}\bar{\Sigma}^{0}$ cross sections at center-of-mass energies from $2.3864$ to $3.0200$\gev}
\author{
M.~Ablikim$^{1}$, M.~N.~Achasov$^{10,d}$, P.~Adlarson$^{64}$, S.~Ahmed$^{15}$, M.~Albrecht$^{4}$, A.~Amoroso$^{63A,63C}$, Q.~An$^{60,48}$, ~Anita$^{21}$, Y.~Bai$^{47}$, O.~Bakina$^{29}$
% ...
\\
\vspace{0.2cm}
(BESIII Collaboration)\\
\vspace{0.2cm} 
$^{1}$ Institute of High Energy Physics, Beijing 100049, People's Republic of China\\
% ...
$^{4}$ Bochum Ruhr-University, D-44780 Bochum, Germany\\
% ...
$^{10}$ G.I. Budker Institute of Nuclear Physics SB RAS (BINP), Novosibirsk 630090, Russia\\
% ...
$^{15}$ Helmholtz Institute Mainz, Johann-Joachim-Becher-Weg 45, D-55099 Mainz, Germany\\
% ...
$^{21}$ Indian Institute of Technology Madras, Chennai 600036, India\\
% ...
$^{29}$ Joint Institute for Nuclear Research, 141980 Dubna, Moscow region, Russia\\
% ...
$^{47}$ Southeast University, Nanjing 211100, People's Republic of China\\
$^{48}$ State Key Laboratory of Particle Detection and Electronics, Beijing 100049, Hefei 230026, People's Republic of China\\
% ...
$^{60}$ University of Science and Technology of China, Hefei 230026, People's Republic of China\\
% ...
$^{63}$ (A)University of Turin, I-10125, Turin, Italy; (B)University of Eastern Piedmont, I-15121, Alessandria, Italy; (C)INFN, I-10125, Turin, Italy\\
$^{64}$ Uppsala University, Box 516, SE-75120 Uppsala, Sweden\\
% ...
$^{d}$ Also at the Novosibirsk State University, Novosibirsk, 630090, Russia\\
% ...
\vspace{0.4cm}
}
\date{\today}

\begin{abstract}
The Born cross sections of  $e^{+}e^{-}\to \Sigma^{0}\bar{\Sigma}^{0}$ are measured at center-of-mass energies from $2.3864$ to $3.0200$~GeV
using data samples with an integrated luminosity of $328.5$~pb$^{-1}$
collected with the BESIII detector operating at the BEPCII collider.  The analysis makes use of a novel reconstruction method for energies near production threshold, while 
a single-tag method is employed at other center-of-mass energies. The measured cross sections are consistent with earlier results from BaBar, with a substantially improved precision. The cross-section lineshape can be well described by a perturbative QCD-driven energy function. In addition, the effective form factors of the $\sig$ baryon are determined. The results provide  precise experimental input for testing various theoretical predictions.

\end{abstract}
\begin{keyword}
BESIII, $\Sigma^{0}$ hyperon \sep Born cross section \sep Form factor
\end{keyword}
\end{frontmatter}
\begin{multicols}{2}
\section{Introduction}
Studies to gain a better understanding of 
 the internal structure of the nucleons have been ongoing ever since the discovery of their non-pointlike nature~\cite{hofstadter}. 
As discussed in Refs.~\cite{chiral, pqcd, lattice1}, the electromagnetic form factors~(EMFFs) are fundamental nucleon observables that are closely related to their internal structure and dynamics.
Further insight into  nucleon structure can also be obtained by studying hyperons that contain one or more strange quarks~\cite{hyperon1,hyperon2}.

Due to their instability, the EMFFs of hyperons are mostly studied in the time-like region, {\it e.g.} via electron-positron annihilation into
a hyperon-antihyperon pair.
Experimentally, many unexpected features have been observed concerning the reaction $e^{+} e^{-} \rightarrow B \bar{B}$ close to their production thresholds, where $B$ denotes a baryon~\cite{bes_fu}. In the charged-baryon sector, a steep rise in the cross section near the production threshold followed by a plateau is observed for $e^{+} e^{-} \rightarrow p \bar{p}$~\cite{time_babar, time_cmd} and $e^{+} e^{-}\rightarrow \Lambda^{+}_{c} \bar{\Lambda}^{-}_{c}$~\cite{lambdac_bes3}.
In the neutral baryon sectors, a non-vanishing cross section near threshold has been observed for $\ee\to n\bar{n}$~\cite{neutron} and $\ee\to\Lambda \bar{\Lambda}$~\cite{lamlam}. 
These abnormal threshold effects have been extensively discussed in the literature, where they are interpreted as being caused by final-state interactions~\cite{int_fsi}, bound states or unobserved meson resonances~\cite{int_res}, or  attractive and repulsive Coulomb forces among quarks~\cite{int_coulomb}. 
To decide between these hypotheses and gain a better understanding of the nature of these effects,  
more measurements are required on baryon-pair production, especially near  production threshold.

Recently, the reactions  $\ee\to\Sigma^{\pm}\bar{\Sigma}^{\mp}$~\cite{sigma} and $\ee\to\Xi\bar{\Xi}$~\cite{cascas} have been investigated
at BESIII with an energy-scan approach. No abnormal threshold effects are observed in these processes. Instead, a perturbative-QCD driven function can describe the cross section lineshapes well. In addition, it is found that the cross sections for $\ee\to\Sigma^{-}\bar{\Sigma}^{+}$ are consistently smaller than those for $\ee\to\Sigma^{+}\bar{\Sigma}^{-}$, and the corresponding EMFFs of $\Sigma^{\pm}$ are proportional to the incoherent sum of the squared charges of their valence quarks, 
$\sum_{q} Q_{q}^{2}$ with $q=u,d,s$ quarks. Thus, it is interesting to measure the Born cross section of $\ee\to\Sigma^{0}\bar{\Sigma}^{0}$ near the production threshold and validate whether the same asymmetry appears in the EMFFs of $\Sigma^{0}$ in the same energy region. In addition, the study of $\ee\to\Sigma^{0}\bar{\Sigma}^{0}$ provides  important experimental input
for testing the diquark correlation model~\cite{di}, which predicts that the cross section of $\ee\to\sigg$ reaction is significantly suppressed compared to that of $\ee\to\lambb$. Previous measurements of $\ee\to\Sigma^{0}\bar{\Sigma}^{0}$ have been performed at BaBar via an initial-state radiation~(ISR) approach~\cite{babar}. However, the uncertainty on these measurements  is too large to address the above questions. Thus, it is important to study the $\ee\to\Sigma^{0}\bar{\Sigma}^{0}$ reaction at collision energies at and above its production threshold with improved precision.
 
In this Letter, we present a measurement of the process $e^{+}e^{-}\to\Sigma^{0}\bar{\Sigma}^{0}$ at BESIII, using data samples
corresponding to an integrated luminosity of $328.5$~pb$^{-1}$ at center-of-mass (c.m.)~energies ranging from $2.3864~\rm{to}~3.0200\gev$~\cite{dataset}. A novel method is applied to reconstruct the secondary particles from antiproton interactions for signal process near production threshold~\cite{lamlam}, and a single-tag method that reconstructs $\Sigma^{0}\to\gamma\Lambda$ and $\Lambda\to p\pi^{+}$ is performed for other c.m.~energies. The charge-conjugate state of the $\Sigma^{0}$ mode is always implied when discussing the single-tag method.

\section{Detector and Monte Carlo simulation}
The BESIII detector is a magnetic spectrometer~\cite{Ablikim:2009aa} located at Beijing Electron Positron Collider (BEPCII).  The cylindrical core of the BESIII detector covers $93\%$ of the full solid angle and consists of a helium-based multilayer drift chamber (MDC), a plastic scintillator time-of-flight system (TOF), and a $\mathrm{CsI}(\mathrm{Tl})$ electromagnetic calorimeter (EMC). The subdetectors are enclosed in a superconducting solenoid magnet with a field strength of $1.0~\mathrm{T}$. The solenoid is supported by an octagonal flux-return yoke with resistive plate counter muon-identifier modules interleaved with steel. The momentum resolution of charged particles is $0.5\% $ at $1\gev$, and the energy loss (d$E$/d$x$) measurement provided by the MDC has a resolution of $6\%$ for the electrons from Bhabha scattering. The energy resolution for the photons is $2.5\% ~(5\%) $ at $1\gev$ in the barrel (end cap) of the EMC. The time resolution of the TOF is $68~\mathrm{ps}$ $(110~\mathrm{ps})$ in the barrel (end-cap) region. 

Simulated Monte Carlo~(MC) samples generated with {\sc geant4}-based~\cite{geant4} package, which includes the geometric description of the BESIII detector, and the detector response, are used to determine detection efficiency and to estimate background contributions. In this analysis, the generator software package {\sc conexc}~\cite{conexc} is used to simulate the signal MC samples  $e^{+}e^{-}\to\Sigma^{0}\bar{\Sigma}^{0}$,
and calculate the corresponding correction factors for higher-order processes with one radiative photon. The angular amplitude of this process  is 
taken to be uniform in phase space~(PHSP) at $\sqrt{s}=2.3864$~GeV due to
the by-definition equality of electric~$(G_{E})$ and magnetic ($G_{M}$) FFs at threshold,
and is distributed according to the  differential amplitude presented in Ref.~\cite{Anjrej} at other c.m.~energies. The studies of  $e^{+}e^{-}\to\Sigma^{0}\bar{\Sigma}^{0}$ near threshold employ a sample in which the subsequent decay modes of both $\Sigma^{0}$ and $\bar{\Sigma}^{0}$ are simulated using {\sc evtgen}~\cite{bes} with an exclusive decay chain $e^{+}e^{-}\to\Sigma^{0}\bar{\Sigma}^{0}\to\gamma\gamma\Lambda\bar{\Lambda}\to\gamma\gamma p\bar{p}\pi^{+}\pi^{-}$. For the higher energy studies of $e^{+}e^{-}\to\Sigma^{0}\bar{\Sigma}^{0}$ with a single-tag method, the decay mode of $\Sigma^{0}\to\gamma\Lambda$ with subsequent  decay $\Lambda\to p\pi^{-}$ is simulated, while the $\bar{\Sigma}^{0}$ is allowed to decay inclusively according to the branching fractions reported by the  Particle Data Group~(PDG)~\cite{pdg}. Large simulated samples of generic $e^{+} e^{-} \rightarrow$~hadrons events implemented by the {\sc LundArlw} generator~\cite{lundarlw} are used to estimate possible backgrounds. Backgrounds coming from the QED processes $e^{+}e^{-}\to l^{+}l^{-}~(l=e,\mu)$ and $e^{+}e^{-}\to\gamma\gamma$ are investigated  with {\sc babayaga}~\cite{babayaga},  while {\sc bestwogam}~\cite{bestwogam} is used for two-photon processes. The dominant background channels, $e^{+}e^{-}\to\Lambda\bar{\Lambda}\to p\bar{p}\pi^{+}\pi^{-}$ and $e^{+}e^{-}\to\Lambda\bar{\Sigma}^{0}+c.c$, are generated exclusively using the phase-space {\sc conexc}~\cite{conexc} generator. The $\Sigma^{0}$ and $\Lambda$ particles are simulated in the $\sig\to\gamma\Lambda$ and $\lam\to p\pi^{-}$ decay modes.


\section{Formalism}

Under the one-photon exchange approximation, the Born cross section for the process $e^{+} e^{-} \to B \bar{B}$, where $B$ is a spin-1/2 baryon, can be expressed in terms of $G_{E}$ and $G_{M}$ FFs~\cite{ee} as
\begin{equation}
	\label{equ:eq1}
	\sigma^{\text{B}}(s)=\frac{4 \pi \alpha^{2} \beta C}{3 s} \left[\left|G_{M}(s)\right|^{2}+\frac{1}{2 \tau}\left|G_{E}(s)\right|^{2}\right].
\end{equation}
Here, $\alpha$ is the fine-structure constant, $s$ is the square of the c.m.~energy, $\beta=\sqrt{1-1/\tau}$ is a phase-space factor, $\tau= s/4m_{B}^{2}$, $m_{B}$ is the baryon mass and $C$ is the Coulomb enhancement factor that accounts for the $B\bar{B}$ final-state interaction~\cite{schwinger,coulomb}. The factor $C$ is equal to unity for pairs of neutral baryons and $y/(1-e^{-y})$ with $y=\pi\alpha(1+\beta^{2})/\beta$ for pairs of charged baryons. 
From Eq.~(\ref{equ:eq1}), it follows that the Born cross section 
of $e^{+} e^{-} \rightarrow B \bar{B}$ is non-zero at the production threshold ($\beta \rightarrow 0$) for charged baryon pairs, while it should vanish at the threshold for neutral baryon pairs. 


The magnitude of the effective form factor $|G_{\text{eff}}|$ is defined by the combination of the $G_{E}$ and $G_{M}$ FFs~\cite{hyperon_babar}~as

\begin{equation}
\label{eq:geff}
	\left|G_{\text{eff}}(s)\right|\equiv\sqrt{\frac{2 \tau\left|G_{M}(s)\right|^{2}+\left|G_{E}(s)\right|^{2}}{2 \tau+1}}.
\end{equation}
 
Experimentally, the Born cross section of $e^{+}e^{-}\to\Sigma^{0}\bar{\Sigma}^{0}$ can be determined by
\begin{equation}
	\label{eqn:calsigma}
	\sigma^{\text{B}}(s) = \frac{N_{\mathrm{obs}}}{\mathcal{L}(1+\delta^{r})\frac{1}{|1-\Pi|^{2}}~\varepsilon~\mathcal{B}},
\end{equation} 
where $N_{\mathrm{obs}}$ is the yield of signal events in data, $\mathcal{L}$ is the integrated luminosity, $(1+\delta^{r})$ is the ISR correction factor.
$\frac{1}{|1-\Pi|^{2}}$ is the vacuum polarization factor~\cite{VP}, and $\varepsilon$ is the detection efficiency determined from simulated MC events.
The factor $\mathcal{B}$ is the product of the relevant daughter branching fractions, i.e.,
 $\mathcal{B}=\mathcal{B}(\Lambda \rightarrow p \pi^{-})\cdot\mathcal{B}(\bar{\Lambda} \rightarrow \bar{p} \pi^{+})$ for exclusive MC near production threshold,
and  $\mathcal{B}=\mathcal{B}(\Lambda \rightarrow p \pi^{-})$ for semi-inclusive MC at other c.m.~energies.
We note that as the branching fraction of the decay $\Sigma^{0}\to\gamma\Lambda$ is $100\%$, it can be omitted from the expression for  $\mathcal{B}$~\cite{pdg}.

With the Born cross section obtained experimentally, $|G_{\text{eff}}|$ can be determined through substitution of Eq.~(\ref{equ:eq1}) into Eq.~(\ref{eq:geff}) to yield

\begin{equation}
	\label{eqn:fac}
	\left|G_{\text {eff }}(s)\right|=\sqrt{\frac{\sigma(s)}{\frac{4 \pi \alpha^{2} \beta C}{3s}\left[1+\frac{1}{2 \tau}\right]}},
\end{equation}
which is proportional  to the square root of the Born cross section.

\section{Data analysis at $\sqrt{s}=2.3864~\mathrm{\bf and}~2.3960\gev$}
\label{sec:thres}

In this section, the process $\ee\to\sigg$~at~$\sqrt{s}=2.3864 \gev$, which is approximately $1.2\mev$ above production threshold, and at $\sqrt{s}= 2.3960 \gev$ is selected with the final state topology $\gamma\gamma p \bar{p} \pi^{+} \pi^{-}$. Due to the low momenta of final state particles near threshold and the small PHSP in $\sig(\sigb)$ decays, the proton and antiproton in the final states are not easily detected in the MDC. Furthermore, the detection efficiency of signal photons is low at these energies, and there is a high multiplicity of background photons.  Thus, we do not search for the protons and the photons from the signal channel. Instead, we reconstruct the soft pions and secondary products of antiproton interactions, following the method described in Ref.~\cite{lamlam}. 

Although the final-state pions from the signal channel have a low momentum, and, thereby, a low detection efficiency, these charged particles can be still detected by the MDC. A charged track must fulfill the condition $|\!\cos\theta|<0.93$, where $\theta$ is the polar angle with respect to the direction of the positron beam. The closest approach to the interaction point (IP) is required to be less than $10$~cm along the beam direction ($\left|V_{{z}}\right|<10\,\text{cm}$), and less than $1$~cm in the plane perpendicular to the beam ($V_{xy}<1~\text{cm}$). If a charged track does not satisfy the above requirements, it is considered to be a not good (NG) charged track. Only the d$E$/d$x$ information obtained from the MDC is used to compute particle identification (PID) confidence levels for a pion, kaon and proton hypotheses, as the low momentum of the particles under consideration prevents them reaching the TOF system. Each charged track is assigned to the particle type with the highest confidence level. The candidate events are required to have two charged tracks identified as $\pi^{+}~\textrm{and}\,~\pi^{-}$, associated with the pions from the $\Lambda (\lamb)$ decay. The two charged pion tracks are constrained to a common decay vertex by applying a vertex fit. The transverse distance of the decay vertex from the IP is required to be less than $2~\mathrm{cm}$. 

\begin{figure}[H]
	\vspace{-1.35cm}
	\begin{center}
		\begin{overpic}[width=7.0cm,height=5.0cm,angle=0]{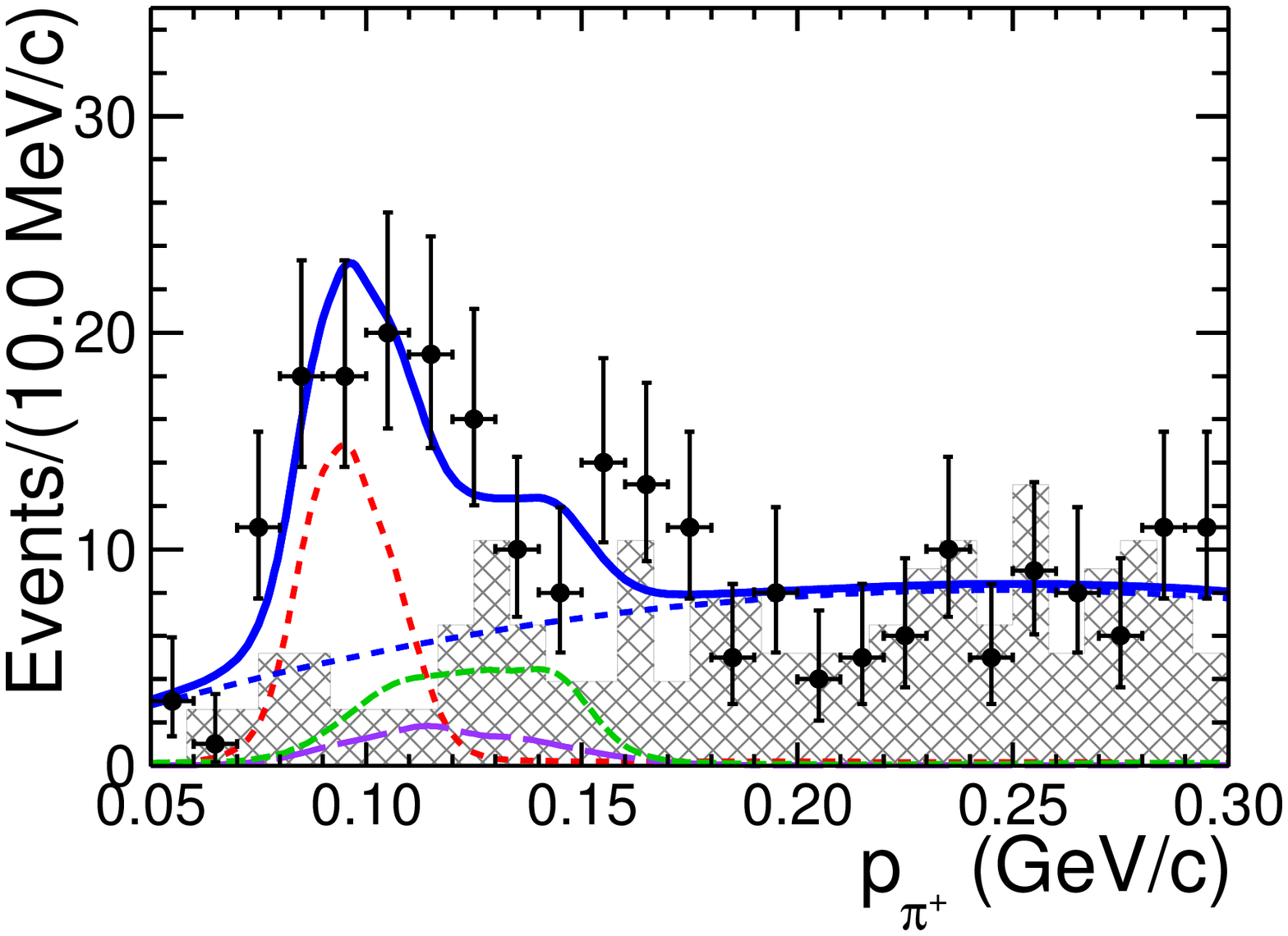}
			\put(23,60){\bf\small{(b)}}
		\end{overpic}
		\vspace{-0.3cm}
		\caption{Momentum distributions of reconstructed $\pi^{+}$ tracks for selected $e^{+}e^{-}\to \sigg$ events at (a) $\sqrt{s}=2.3864$~GeV and (b) $\sqrt{s}=2.3960$~GeV. Dots with error bars (in black) are from data; hatched histograms (in grey) are the combinatorial background events of normalized beam-related background and generic hadronic MC samples after removing the peaking exclusive background channels. The solid blue curve represents the total fit, red dashed curve the signal, dashed blue curve the non-peaking backgrounds, whereas the long dashed violet curve and dashed green curve are the fixed background lineshapes from the exclusive MC samples of $\ee\to\Lambda\bar{\Lambda}$ and $\ee\to\Lambda\bar{\Sigma}^{0}+c.c.$, respectively.}
		\vspace{-0.5cm}
		\label{inc_ana}
	\end{center}
\end{figure}

The antiproton from the signal decay interacts with the nucleons of the detector material, mostly in the beam-pipe, and produces several secondary particles. To identify the secondary particles produced from the antiproton, at least two NG tracks are required. A vertex fit is then applied to these secondary tracks. Furthermore, the transverse distance of the decay vertex from the beam pipe is required to lie within $(1,5)$~cm. This range is set following the results of a signal MC study which shows the majority of decay vertices to be at around $3$~cm, which is the radius of beam-pipe. It is further required that $\theta_{\pi^{+}\pi^{-}}\in(20^{\circ},170^{\circ})$, where  $\theta_{\pi^{+}\pi^{-}}$ is the opening angle between the $\pip$ and the $\pin$, to remove cosmic rays and $\gamma$-conversion ($\gamma\to\ee$) events. To suppress the combinitorial background, the momentum of another $\pi^{-}$ tracks is required to be located in the signal region $(0.08, 0.12)\gevc$. The momentum of $\pi^{+}$ tracks is shown in Fig.~\ref{inc_ana}.

Potential sources of residual contamination that survive the selection are investigated by studying the generic hadronic MC samples with an event type analysis tool, TopoAna~\cite{ana_tpo}. 
The main peaking background originates from $\ee\to\lambb$  and $\ee\to \bar{\Lambda}\Sigma^{0}+c.c.$, which are decays with one or two fewer photons than the signal final state. 
For these processes, dedicated exclusive MC samples are generated to estimate their contributions in the signal region. 
The contribution from beam-associated background events is estimated from a sample of data collected with separated beams  at~$\sqrt{s}$~$=$~$2.6444$~GeV.  
Detailed MC studies indicate that the generic hadronic MC samples are distributed smoothly after removing the above exclusive background channels, as well as the beam-associated background in the region of interest, as illustrated in Fig.~\ref{inc_ana}.

\begin{figure}[H]
	\vspace{-0.3cm}
	\begin{center}
		\begin{overpic}[width=7.0cm,height=5.0cm,angle=0]{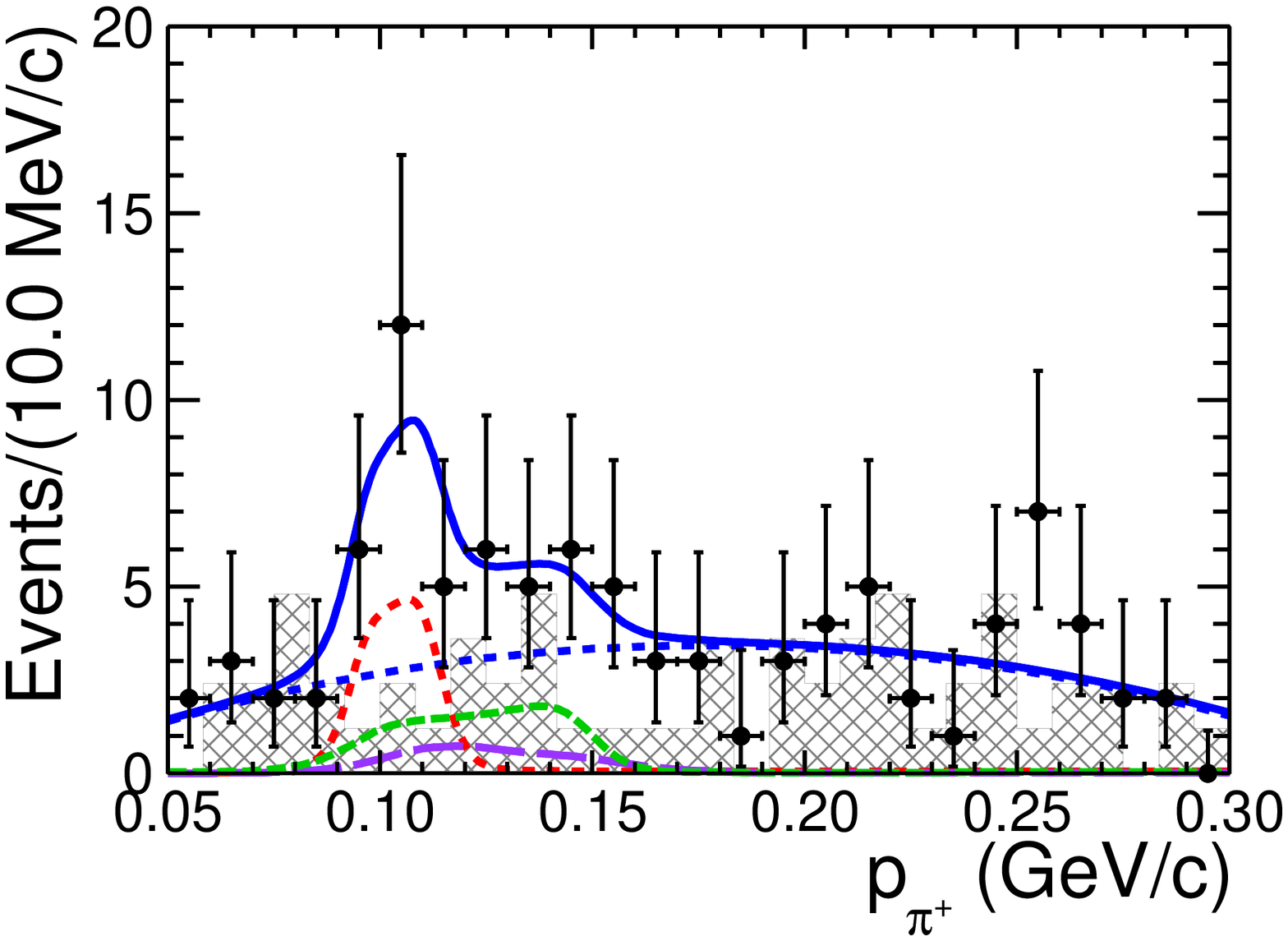}
			\put(23,60){\bf\small{(a)}}
		\end{overpic}
	\end{center}
\end{figure}

To determine the signal yields, an unbinned maximum-likelihood fit is performed to the momentum of  $\pip$ tracks ($p_{\pi^{+}}$). The signal probability-density function (PDF) is described with a MC-simulated shape convolved with a Gaussian function to compensate for possible mass-resolution differences between data and MC simulation. 
The backgrounds from $\ee\to\lam\lamb$ and $\ee\to\bar{\Lambda}\Sigma^{0}+c.c.$  are 
described by the MC-determined shapes, where the yields normalized according to the luminosity and their cross sections~\cite{lamlam} are fixed in the fit.
The PDF for other generic hadronic MC samples and beam-associated backgrounds are described by a second-order polynomial function. The fit results at $\sqrt{s}=$~$2.3864$ and $2.3960$~GeV are shown in Fig.~\ref{inc_ana}. The statistical significances of the signals in $p_{\pip}$ distributions at $\sqrt{s}= 2.3864$ and $2.3960 \gev$ are found to be $2.3\sigma$ and $4.5\sigma$, respectively, which are
determined according to Wilk's theorem~\cite{wilk} from the differences of the log-likelihood values and degrees of freedom with and without considering the signal process in the fits. 
As a cross-check, an alternative two-dimensional fit is performed to the momentum of $\pip$ and $\pin$ tracks at each energy point, which is found to give compatible results.
The Born cross sections of  $e^{+}e^{-}\to\Sigma^{0}\bar{\Sigma}^{0}$ and effective FFs of $\Sigma^{0}$ 
at  $\sqrt{s}=2.3864$ and $2.396$~GeV can be determined using Eq.~(\ref{eqn:calsigma}) and Eq.~(\ref{eqn:fac}).
The results and the inputs used in the calculation are 
summarized in Table~\ref{born_cross}.

Since the statistical significance of signal events is less than 3$\sigma$ at $2.3864$~GeV, the upper limit on the number of signal events ($N_{\rm{U.L.}}$) is estimated at the $90\%$ confidence level (C.L.) using a Bayesian method~\cite{bays}. We separate the systematic uncertainties into two categories: correlated uncertainties associated with the assumed signal and background shapes and fit range, and the remainder that are considered uncorrelated.
To take into account the uncorrelated uncertainties related to the fit procedure, two alternative fit scenarios are considered: $(1)$ changing the momentum range by $5 \mevc$; and $(2)$ replacing the second-order polynomial function with the exact background distribution seen in Fig.~\ref{inc_ana}. We consider all combinations of the two scenarios and select the likelihood curve with the maximum $N_{\rm{U.L.}}$ as a conservative estimate. The selected likelihood distribution is then smeared by the remaining so-called correlated uncertainties, whose components are described below. The calculation of the U.L on the Born cross section and effective FF at the $90\%$ C.L. is performed analogously by replacing the $N_{\rm{obs}}$ with $N_{\rm{U.L.}}$, as shown in Table~\ref{born_cross}. 

Several sources of systematic uncertainties are considered in the measurement of the Born cross section near the production threshold. 
The $\pi^{\pm}$ tracking (PID) efficiencies are studied using a control sample of $\jpsi\to p\bar{p}\pip\pin$ events. The corresponding systematic uncertainties are estimated as $4.6\%$ ($2.0\%$). The uncertainty of the selection efficiency for antiprotons is estimated to be $0.3\%$~\cite{lamlam}. Alternative fits are performed to study the uncertainty from the fit procedure.  These include varying the fitting range, varying the signal shape by fixing the resolution of the convolved Gaussian to be $\pm1\sigma$ different from its nominal value, changing the background PDF from a second order polynomial function to the background shape from the generic hadronic MC samples combined with the contribution from the beam-related background and varying the cross section of fixed backgrounds lineshapes within the uncertainty of normalized backgrounds events. 

Possible biases due to the various rejection windows are assessed by varying the criteria above and below the nominal selection, as described in Ref.~\cite{sys}. For each systematic variation, the parameters values are re-obtained, $x_{\mathrm{test}}\pm\sigma_{\mathrm{test}}$ and the changes evaluated compared to the nominal values, $\Delta x =|x_{\mathrm{nom.}}-~x_{\mathrm{test}}|$. Also calculated are the uncorrelated uncertainties, $\sigma_{\mathrm{uc.}}=\sqrt{|\sigma^{2}_{\mathrm{nom.}}-\sigma^{2}_{\mathrm{test}}|}$, where $\sigma_{\mathrm{nom.}}$ and $\sigma_{\mathrm{test}}$ correspond to the fit uncertainties of the  nominal and systematic test results, respectively. A systematic uncertainty is assigned, if the ratio $r=\Delta x/\sigma_{\mathrm{uc.}}$ shows a trending behavior and larger than two. 
The effect of the $\theta_{\pip\pin}$ veto is tested by varying the boundaries of the rejection windows from $16^{\circ}$ to  $24^{\circ}$ (lower side) and $166^{\circ}$ to $174^{\circ}$ (higher side) with respect to the nominal selection. Since the corresponding normalised shift is $r<2$, no associated uncertainty is assigned. The uncertainty due to the cross section lineshape is negligible as well, as is discussed below in Sec.~\ref{sec:other}. The integrated luminosity is determined with large angle Bhabha events with an uncertainty of $1.0\%$~\cite{dataset}. All these systematic uncertainties are treated as uncorrelated and summed in quadrature, giving a total uncertainty of $9.0\%$ and $9.9\%$ for the Born cross section at $\sqrt{s}=2.3864$ and $2.3960$~GeV, respectively. As the effective FF is proportional to the square root of the Born cross section, the corresponding systematic uncertainty on this quantity is half that of the uncertainty on the Born cross section.

\section{Data analysis from $\sqrt{s}=2.5000\,\rm{\bf to}\,3.0200\gev$} 
\label{sec:other}
As the full reconstruction method for selecting the $\ee \to \Sigma^{0}\bar{\Sigma}^{0}$ events has a low reconstruction efficiency, a single-tag $\sig$ baryon technique is employed at energies of $\sqrt{s}=2.5000\,\gev$ and above. We fully reconstruct the $\sig$-prong in the $\gamma\lam$ decay mode with $\Lambda\to p \pi^{-}$. 

Charged tracks are reconstructed within $|\!\cos\theta|<~0.93$ and are required to satisfy $|V_{z}|<30$~cm and $V_{xy} < 10$~cm. The combined information from d$E$/d$x$ and TOF are used to form particle-identification (PID) confidence levels for $\pi,K$ and $p$ hypotheses. Each track is assigned to the particle type corresponding to the highest confidence level. Candidate events with at least two charged tracks identified as a proton and  a pion are kept for further analysis. The photon candidate is required to be within the barrel region $(|\!\cos \theta|<0.8)$ of the EMC with deposited energy of at least $25~ \mathrm{MeV}$, and within the end-cap regions $(0.86<|\!\cos \theta|<0.92)$ with at least $50~\mathrm{MeV}$. In order to suppress electronic noise and showers unrelated to the event, the EMC time difference from the event start time is required to be within ($0, 700$) ns. At least one photon is required in this analysis. 

$\Lambda$ candidates are reconstructed with a vertex fit to all the identified $p \pin$ combinations. A secondary-vertex fit ~\cite{vertex} is then employed for the $\Lambda$ candidate and events are kept if the decay length, $ i.e.$ the distance from the production vertex to the decay vertex, is greater than zero. A mass window of $|M_{p\pi^{-}}- m_{\Lambda}|<6 \mevcc$ is required to select $\lam$ candidates, where $m_{\lam}$ is the known $\lam$ mass from the PDG~\cite{pdg}. In addition, the value of $\chi^2$ from the secondary vertex fit is required to be less than $20$, determined by optimizing the figure of merit $S/\sqrt{S+B}$ based on MC simulation, where $S$ is the number of signal MC events and $B$ is the number of background events in the generic hadronic MC sample normalized according to integrated luminosity and cross section. 
All combinations of the $\Lambda$ and photon candidates are considered, and that combination with the minimum value of 
$|p_{\sig} - p_{\rm exp.}|$ is chosen to determine the photon from $\Sigma^{0}$ decay,
where $p_{\sig}$ and $p_{\rm exp.}$ are the measured and expected momentum of $\Sigma^{0}$, defined as $p_{\rm exp.}\equiv\sqrt{s/4 - m^{2}_{\Sigma^{0}}}$,
with $m_{\Sigma^{0}}$ the mass of $\Sigma^{0}$ from the PDG~\cite{pdg}.
A momentum window $|p_{\sig} -p_{\rm exp.}|<3.5\sigma_{\rm p}$, is then applied to select the $\gamma\lam$ candidate. 
Here both momenta are defined in the c.m.~frame system, and $\sigma_{\rm p}$ is the corresponding optimized momentum resolution, which is about $7\mevc$ at each energy point. 

\begin{figure*}[htbp]
	\begin{center}
		\begin{overpic}[width=5.4cm,height=4.5cm,angle=0]{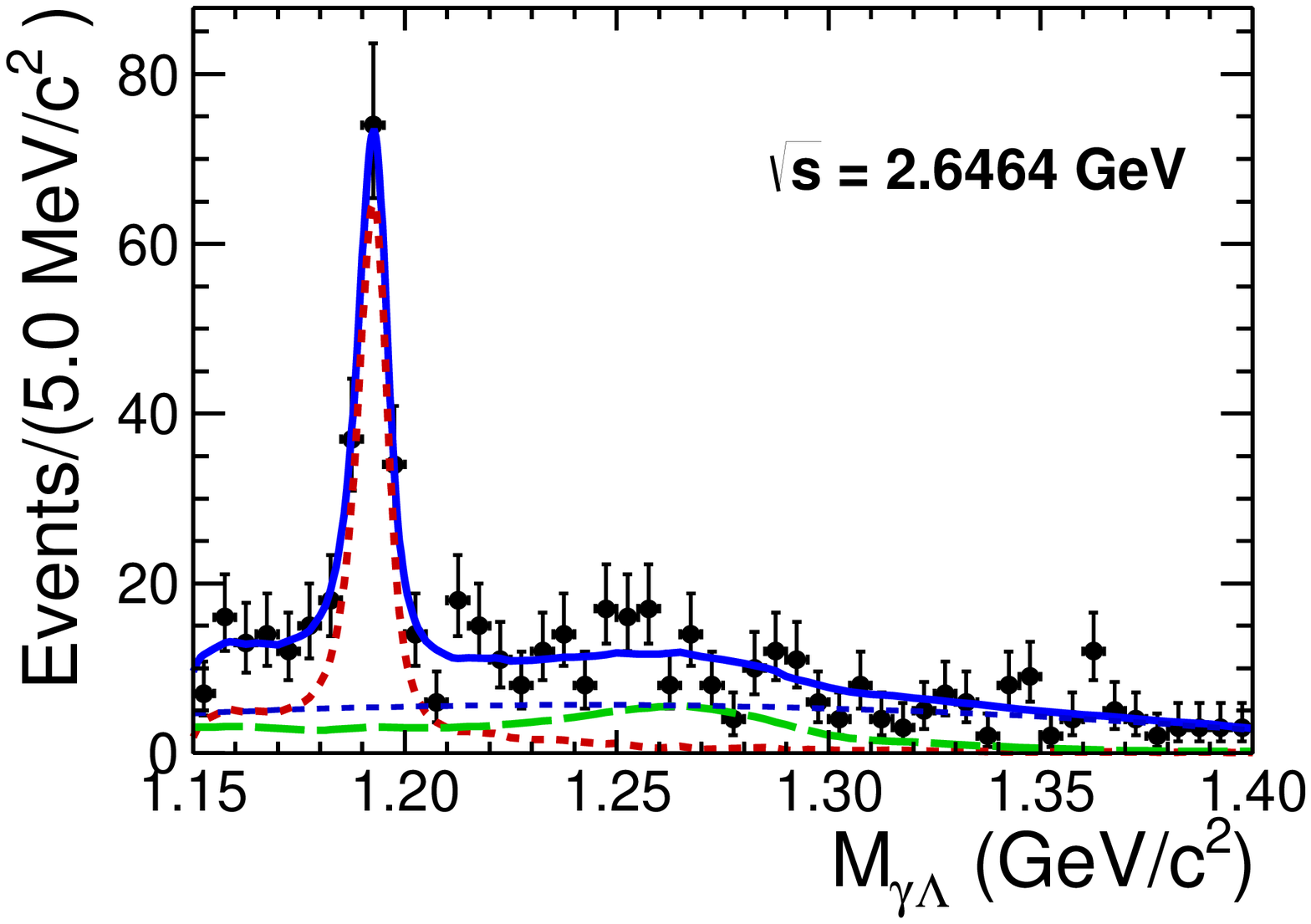}
		\end{overpic}
		\begin{overpic}[width=5.4cm,height=4.5cm,angle=0]{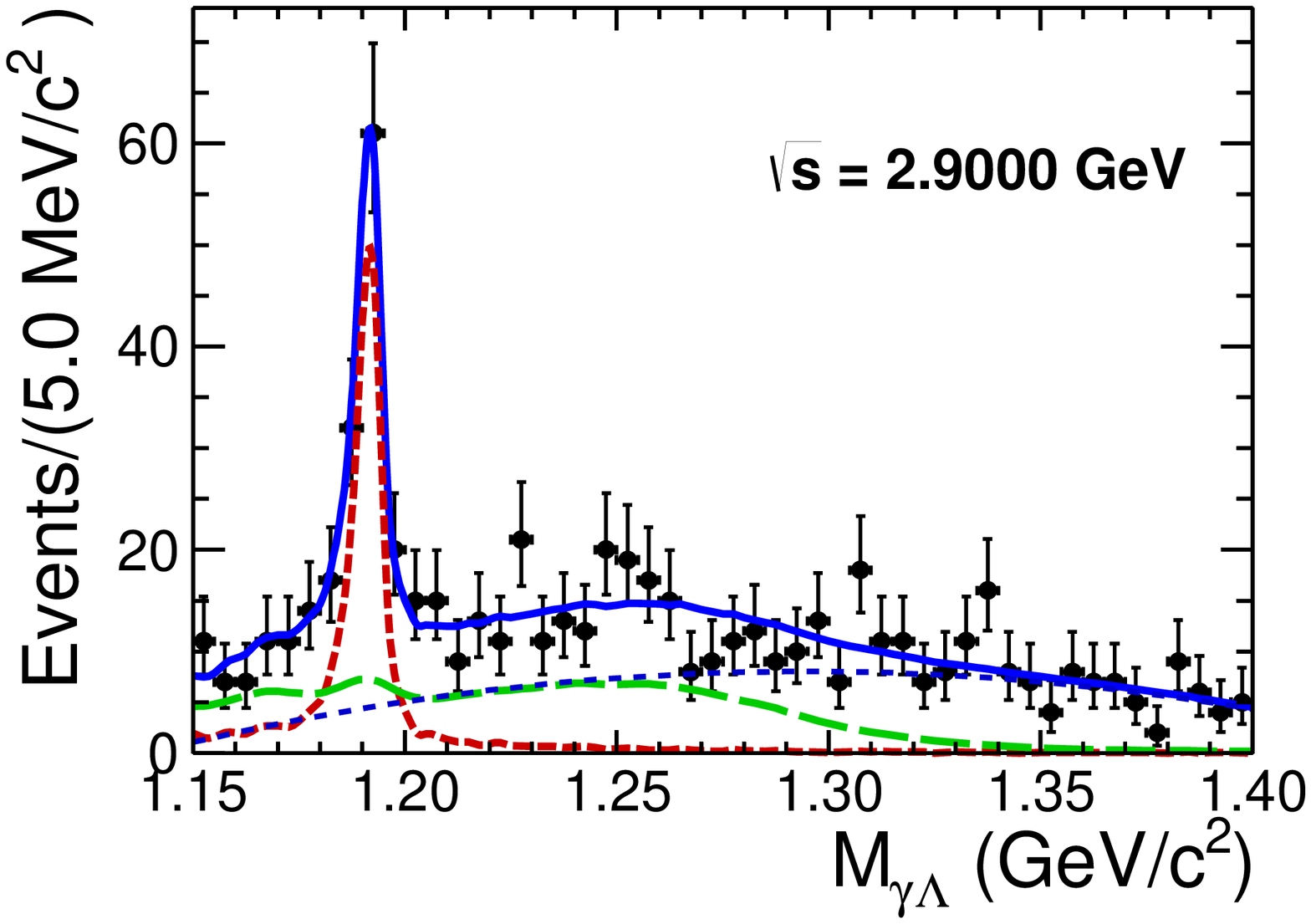}
		\end{overpic}
		\begin{overpic}[width=5.4cm,height=4.5cm,angle=0]{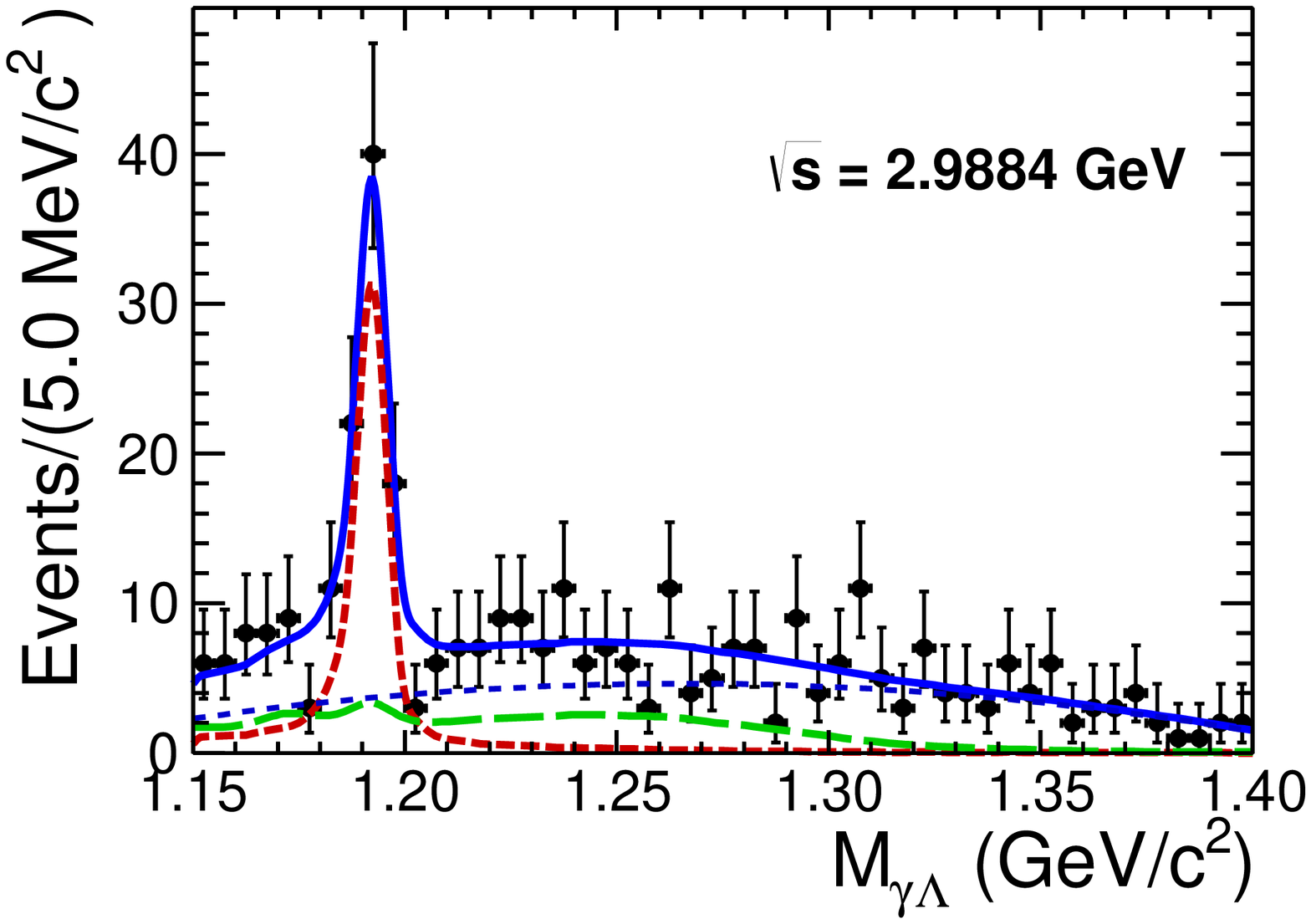}
		\end{overpic}
\caption{ Fit result of the mass spectra $M_{\gamma\lam}$ at each energy point. Dots with error bars are from data, the solid blue curve represents the overall fit, red dashed curve the signal, dashed blue curve the non-peaking backgrounds, and the long dashed green curve is the fixed background lineshape from a MC sample $\ee\to\Lambda\bar{\Sigma}^{0}+c.c.$, respectively.}
\vspace{-0.5cm}
\label{fit_result}
\end{center}
\end{figure*}

The generic hadronic MC samples are used to study possible peaking-background contributions. After applying the same requirements as used for the data, the surviving background is found to originate from $e^+e^-$ annihilation events with the same final state particles as the signal process, with one or more additional $\pi^{0}$, and with one less photon. These background processes are mainly from contributions including intermediate states such as $\lam$, $\Delta$, and $\Sigma$ baryons. For the dominant backgrounds such as $\ee\to\lam\lamb$ and $\ee\to\bar{\Lambda}\Sigma^{0}+c.c.$, dedicated exclusive MC samples are generated to estimate their contributions to the $\gamma \Lambda$ mass spectrum.
A few peaking background events from the process $\ee\to\bar{\Lambda}\Sigma^{0}+c.c.$ contribute to the $\gamma \Lambda$ mass region and their normalized contribution is shown in Fig.~\ref{fit_result}.  There is a negligible contribution from $\ee\to\lam\lamb$ events. 

The signal yields for $\ee\to\sigg$ at each energy point are determined by an unbinned maximum likelihood fit to the $M_{\gamma\lam}$ spectrum. The PDF is described with the MC-simulated shape convolved with a Gaussian function to account for the mass-resolution difference between data and the MC simulation. A second-order polynomial function describes the non-peaking background PDF, whereas the peaking backgrounds are modeled with the MC-determined lineshapes and their respective yields fixed in the fits. Fit results at $\sqrt{s}= 2.3960, 2.5000, 2.6444, 2.6464, 2.9000$ and $2.9884\gev$ are shown in Fig.~\ref{fit_result}. As the single-tag method leads to the double counting effect of the $\sigg$ final-state. It is rather possible that a small fraction of events of both $\Sigma^{0}$ and $\bar{\Sigma}^{0}$ prongs are reconstructed, which introduces the double counting while fitting the $M_{\gamma\Lambda}$ spectrum. A correction factor of $1.09$ for $\sqrt{s}=2.6444$ and $2.6464\gev$ is applied to the statistical uncertainty to account for this effect based on MC simulation~\cite{cascas}, while for other energy points, the correction factor is estimated to be 1.07. The number of observed events are summarized in Table~\ref{born_cross}. The Born cross sections and corresponding effective FFs are determined using Eq.~(\ref{eqn:calsigma}) and Eq.~(\ref{eqn:fac}). The quantities used in the Born cross sections and effective FFs calculations for $\ee\to\sigg$ are summarized in Table~\ref{born_cross}. It should be noted that the data at c.m.~energies $2.9500, 2.9810, 3.0000$ and $3.0200\gev$ are combined to a single luminosity-weighted energy point of  $2.9884\gev$ on account of the limited size of each sub-sample.

\end{multicols}

\begin{figure*}[htbp]
\begin{center}
\vspace{-0.5cm}
\begin{overpic}[width=5.4cm,height=4.5cm,angle=0]{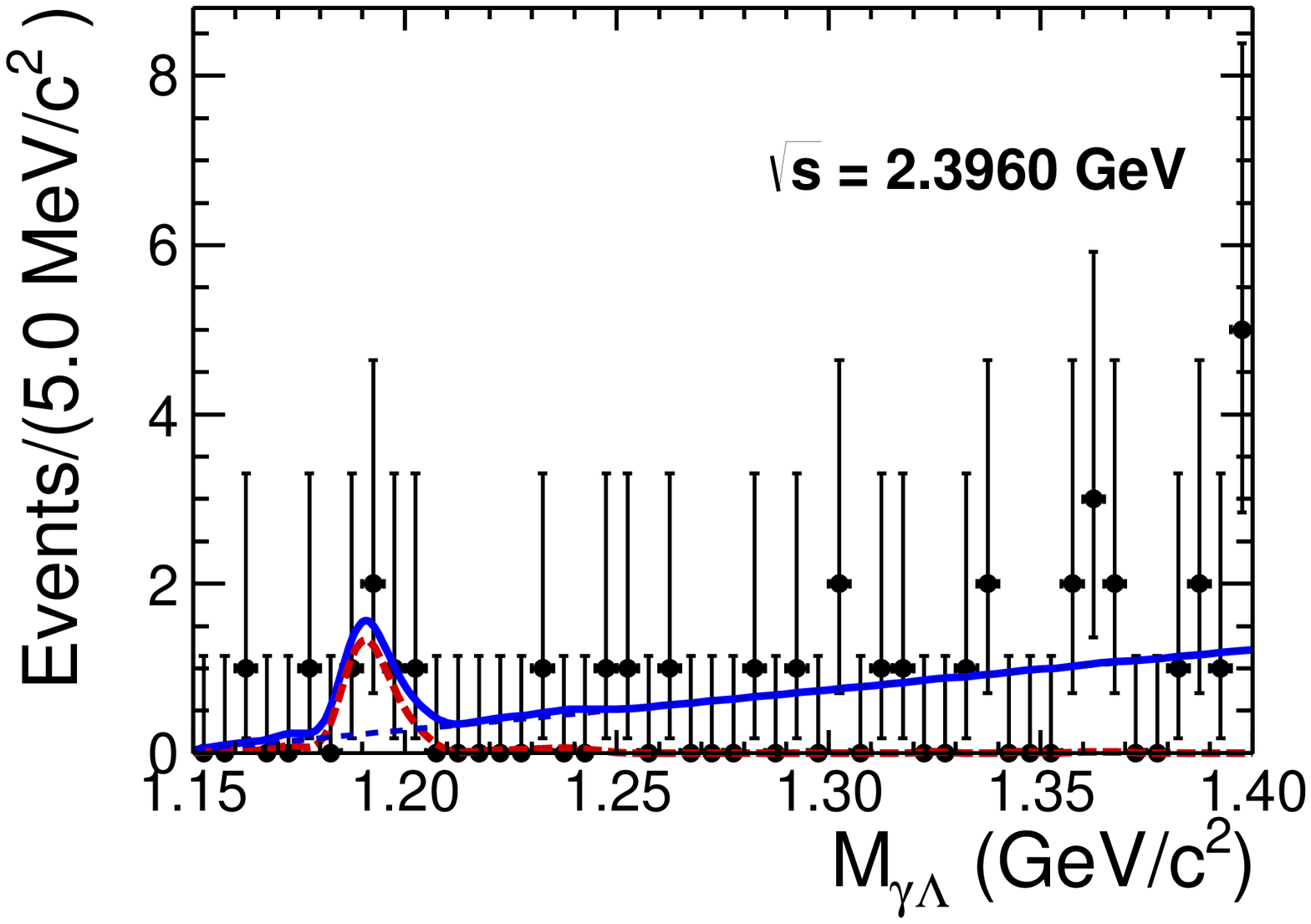}
              \end{overpic}
               \begin{overpic}[width=5.4cm,height=4.5cm,angle=0]{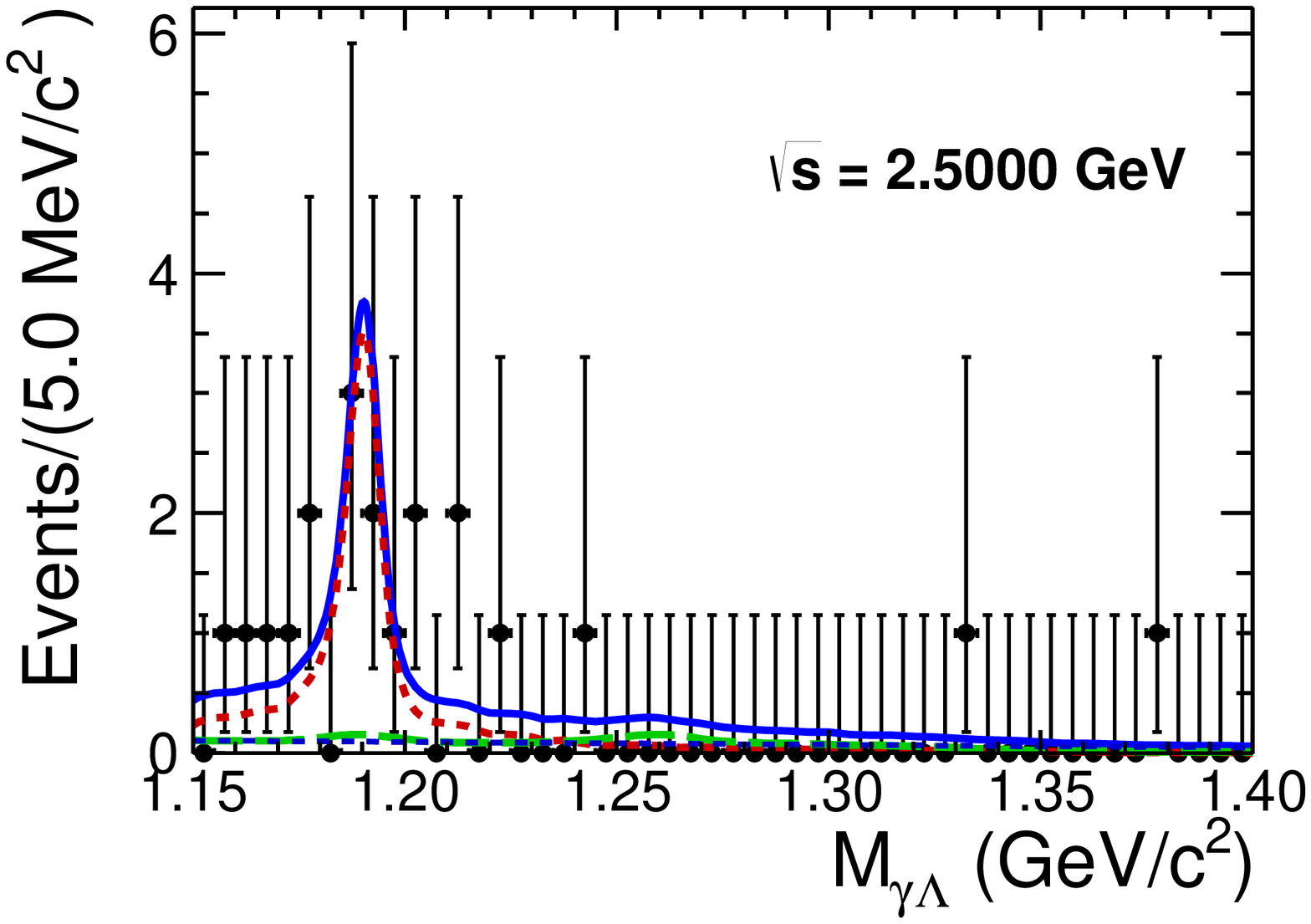}
              \end{overpic}
                \begin{overpic}[width=5.4cm,height=4.5cm,angle=0]{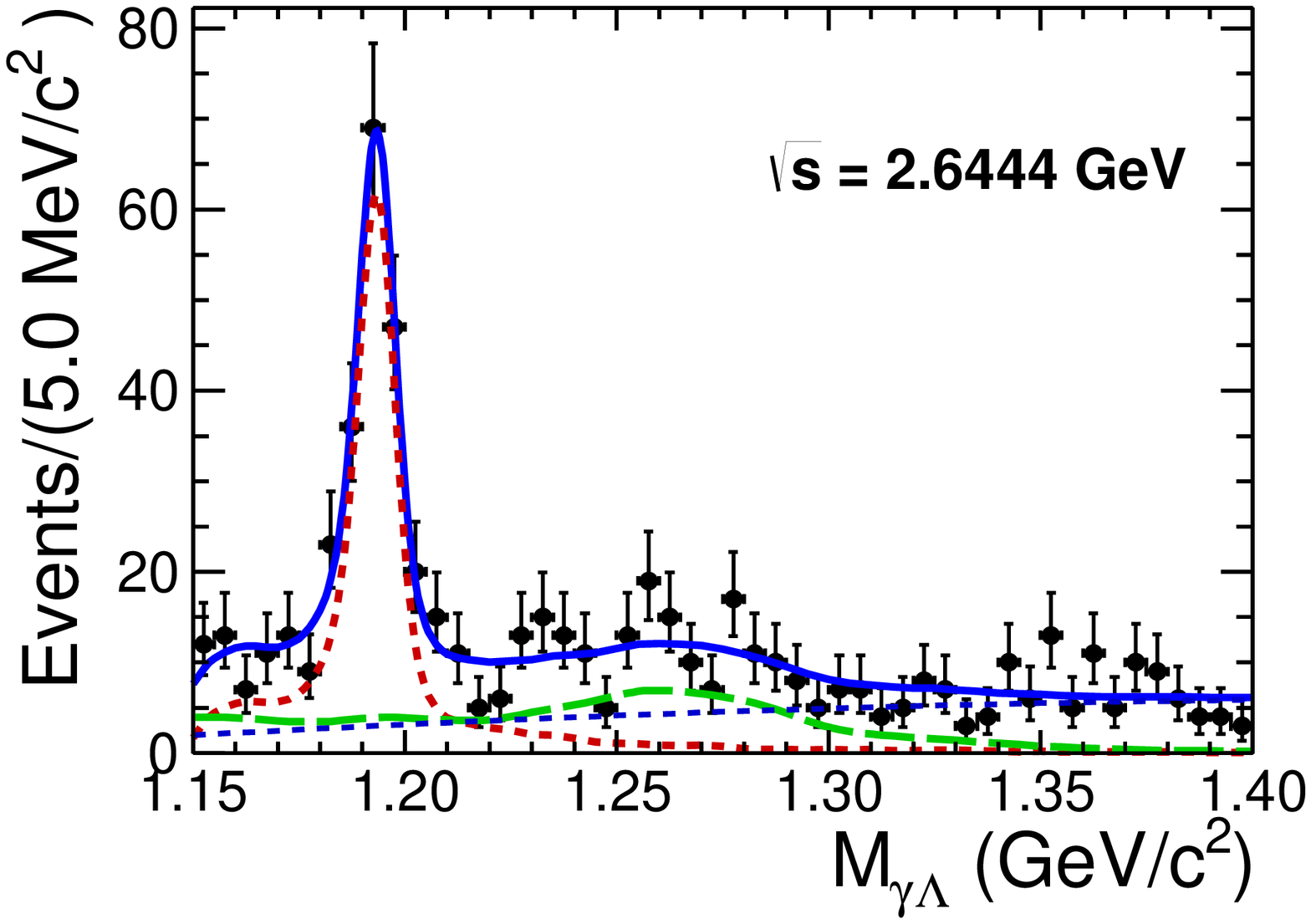}
                \end{overpic}
\vspace{-0.9cm}
\end{center}
\end{figure*}

\begin{multicols}{2}
Sources of systematic uncertainties related to the cross-section measurement include those associated with the $\Lambda$ reconstruction, the photon detection, the momentum window, the fitting method, the cross section lineshape, the angular distribution, the relative phase and the luminosity measurement. The systematic uncertainty on the $\lam$ reconstruction efficiency, associated with the tracking/PID, decay length and mass window requirements is estimated to be $4.5\%$~\citep{lamlam}. The uncertainty on the photon-detection efficiency is assigned by considering a control sample of the decay $J/\psi\to\pip\pin\pi^{0}$ and is found to be $1.0\%$ for each photon~\cite{photon}. The $p_{\sig}$ veto is tested by varying the selection window, $|p_{{\sig}} - {p}_{\rm exp.}|$ from $2\sigma_{\rm p}$ to $6\sigma_{\rm p}$ with the same method as explained above in Sec.~\ref{sec:thres}, and taking  the largest deviation from the nominal result as a conservative estimate of the corresponding uncertainty. Alternative fits are performed to study the uncertainty from the fit procedure. These include varying the fitting range, 
changing the background PDF from a second-order polynomial to a third-order polynomial function and varying the cross section of fixed background lineshape within the uncertainty of normalized background events.  
The uncertainty associated with the description of the signal lineshape is estimated by parameterizing the lineshape in each iteration according to the pQCD power-law function and sampling $500$ lineshapes according to the uncertainty of the parameters and their covariance matrix. The variation in result is found to be much less than $1\%$ and can be neglected. In order to investigate the bias coming from the angular distribution,  the analysis is repeated with the two extreme values of  $G_{E}=0$ and $G_{M}=0$ and the difference in the resulting efficiencies divided by a factor of $\sqrt{12}$ is taken as the uncertainty~\cite{sys}. In addition, to account for effects arising from the uncertainty in the relative phase ($\Delta \Phi$) between the $G_{E}$ and $G_{M}$ FFs, we consider two extreme cases with $\Delta \Phi=0^{\circ}$ and $90^{\circ}$, and find the difference in results to be neglible. The integrated luminosity is determined with large-angle Bhabha events with an uncertainty of $1.0\%$~\cite{dataset}. All the uncertainties are considered uncorrelated and summed in quadrature. They are between $6.1\%$ and $12.0\%$ of the cross sections, depending on the c.m. energy. 
	 
As a cross-check, we compare the Born cross-section result measured at $\sqrt{s}= 2.3960 \gev$ from the single-tag method with that of  the novel approach described in Sec.~\ref{sec:thres}. It is found that both results are consistent within the uncorrelated uncertainty. The result obtained by the novel approach is considered as nominal at this energy point, since it leads to a smaller uncertainty on the cross section than the result obtained using the single-tag method.
\begin{table*}
\centering
\footnotesize 
\caption{The numerical results of  Born cross sections $\sigma^{\mathrm{B}}$ and effective FFs $|G_{\mathrm{eff}}|$ at each c.m.~energy and the quantities used in the calculation, $(1+\delta)= (1+\delta^{r})\frac{1}{|1-\Pi|^{2}}$, defined in the text. The energy point $2.9884$ $\gev$ is the luminosity-weighted combined dataset of the $2.9500, 2.9810, 3.0000$ and $3.0200$ $\gev$ subsamples. The last column shows the statistical significance of the signal $S(\sigma)$ at each energy point. The values in parenthesis represent the corresponding upper limit at the $90\%$ confidence level. The first uncertainty is statistical, and the second is systematic.}
\vspace{-0.1cm}
\label{born_cross}
\begin{tabular}{ccccD{,}{}{4,9}D{,}{}{12,13}D{,}{}{9,12}c}
	\hline \hline
	$\sqrt{s}~$(GeV)
	& $\mathcal{L}(\rm{pb^{-1}})$
	& $ \varepsilon (\%)$ 
	& $1+\delta$
	& N_{\rm{obs}},(N_{\rm{U.L.}})
	&~\sigma^{\rm{B}},(\rm{pb})
	&~|G_{\rm{eff}}|,~(\times 10^{-2})~
	& $S(\sigma)$\\

\hline

$2.3864$   & $22.55$   & $11.1$  & $0.65$ & 11.7 ,~\pm~~5.8~(<28.0)  & 17.7 \pm 8.74  ,\pm 1.59(< 42.4)    & 15.3 \pm 3.78  ,~\pm ~~0.69(< 23.7)   & $2.3$  \\
$2.3960$   &$66.87$    & $7.7$   &$0.75$  &45.1 ,\pm 11.2        &28.6 \pm 7.10  ,\pm 2.75        &11.5 \pm 1.43  ,\pm 0.55        & $4.5$ \\
$2.5000$  &$1.10$       & $32.3$  &$0.94$  &12.7 ,\pm 6.4        &59.6 \pm 30.3 ,\pm 7.15       &9.90  \pm 2.52 ,\pm 0.60       &$3.1$ \\
$2.6444$   &$33.72$    & $47.1$  &$1.10$  &221  ,\pm 25          &19.8 \pm 2.23 ,\pm 1.21      &5.12 \pm 0.29, \pm 0.16      &$12.4$ \\
$2.6464$  &$34.00$    & $46.4 $ &$1.10$  &195  ,\pm 24          &17.6 \pm 2.13 ,\pm 1.20      &4.83 \pm 0.29 ,\pm 0.16       &$11.9$ \\
$2.9000$   &$105.23$   & $40.2$ &$1.44$  &116  ,\pm 17          &2.98 \pm 0.45 ,\pm 0.22      &1.95 \pm 0.15 ,\pm 0.07       &$9.4$ \\
$2.9884$  &$65.18$    & $34.9$  &$1.62$  &78.7 ,\pm 13.9        &3.34 \pm 0.59 ,\pm 0.20     &2.08 \pm 0.18 ,\pm 0.06        &$9.0$ \\ 

\hline 
\hline
\end{tabular}
\end{table*}


\section{Lineshape analysis}

\begin{figure}[H] 
\begin{center}
\begin{overpic}[width=6.5cm,height=5.0cm,angle=0]{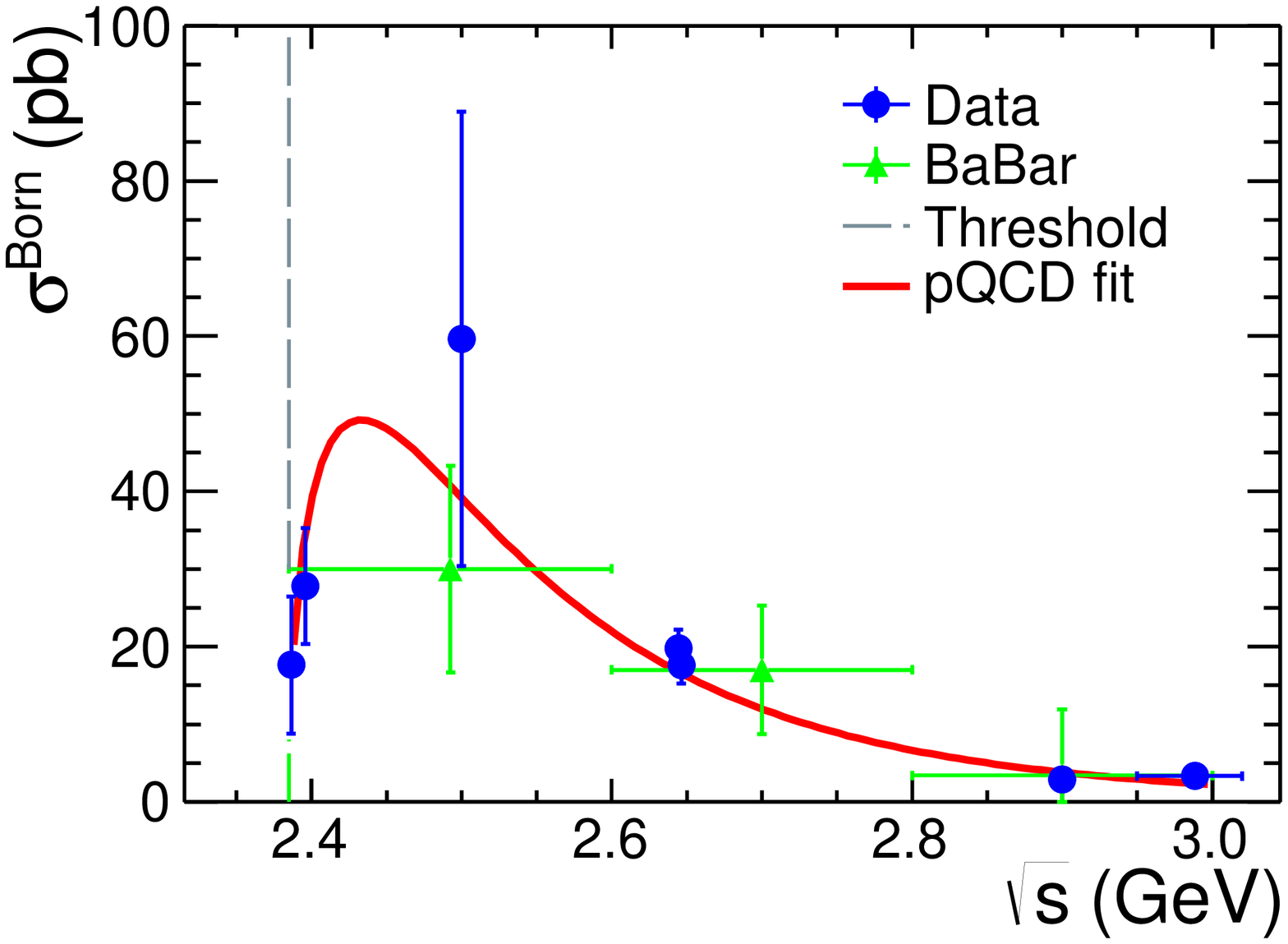}
\put(26,65){\small{\bf (a)}}
\end{overpic}
\hspace{1.5cm}
\begin{overpic}[width=6.5cm,height=5.0cm,angle=0]{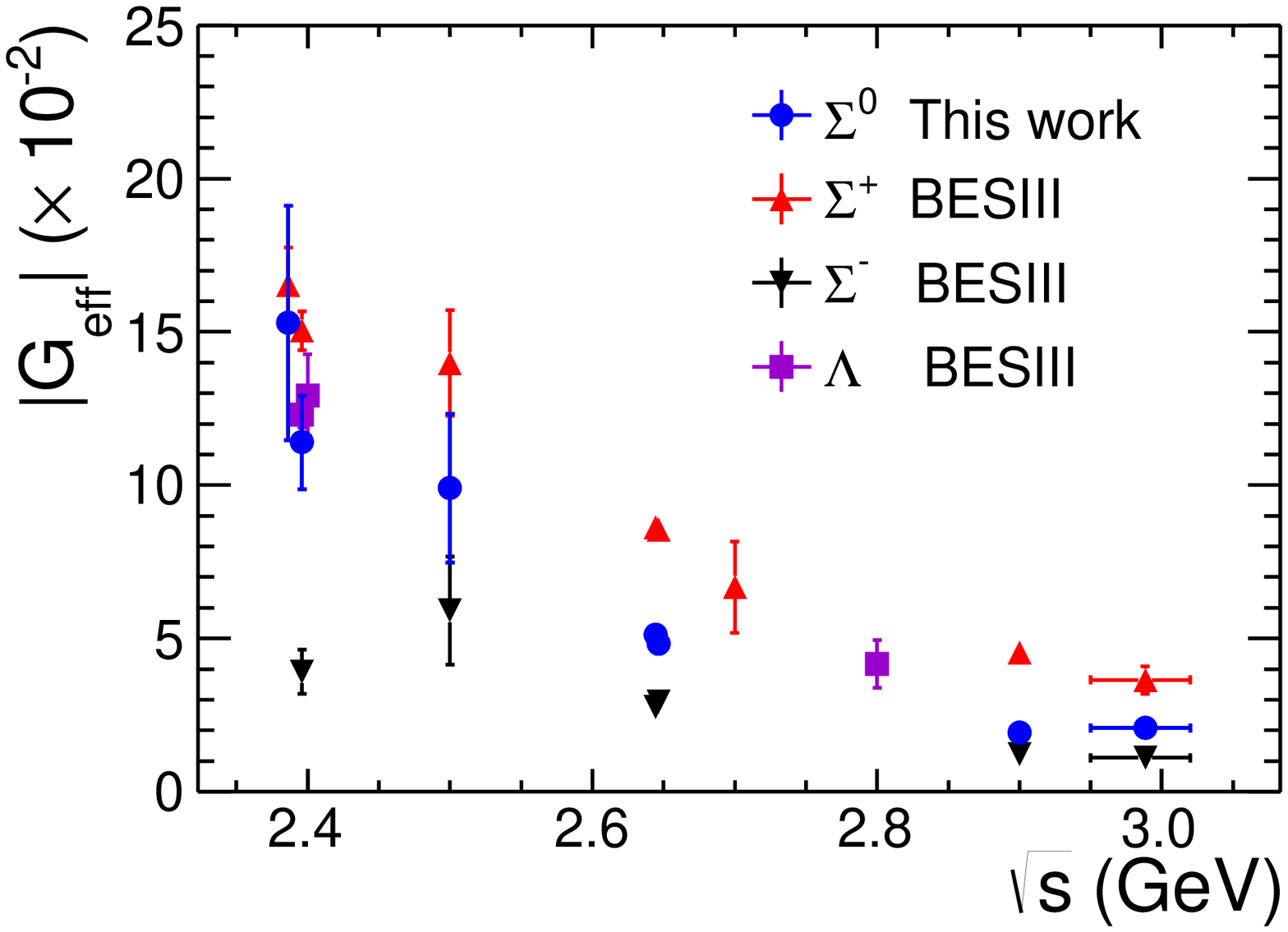}
\put(26,65){\small{\bf (b)}}
\end{overpic}

\caption{(a) Comparison of the cross sections for $e^{+}e^{-}\to\Sigma^{0}\bar{\Sigma}^{0}$ are represented by the dots with error bars in blue from this analysis, while triangles in green are results from BaBar~\cite{babar}. The solid line in red represents the fit with the pQCD energy power function. (b) Comparison of the effective FFs among this work and previous work of BESIII~\cite{sigma, lamlam}. The blue circles represents the results from this analysis. The red upper triangles, black lower triangles and violet squares are from previous BESIII work.}  
\label{comgeff}
\vspace{-0.25cm}
\end{center}
\end{figure}

The measured Born cross-section lineshape of $e^{+}e^{-}\to\sigg$ from $\sqrt{s}=2.3864$ to $3.0200$~GeV is shown in Fig.~\ref{comgeff}~(a).
The cross sections measured in this analysis are in good agreement with that of BaBar, but with improved precision of at least $10\%$ 
at low $\sqrt{s}$ and over $50\%$ above $2.5000\gev$ as depicted in Fig.~\ref{comgeff}~(a). A perturbative QCD-motivated energy power function~\cite{rinaldo}, given by
\begin{equation}
\label{pqcd}
\scalebox{0.78}{$
\begin{aligned}
\sigma^{\text{B}}(s) = \frac{\beta C}{s}\left(1+\frac{2m_{B}^{2}}{s}\right)\frac{c_{0}}{\left(s-c_{1}\right)^{4} \left(\pi^{2}+\text{ln}^{2}\left(s/\Lambda_{\text{QCD}}^{2}\right)\right)^{2}},
\end{aligned}$}
\end{equation}
is used to fit the lineshape, which has been successfully applied in the study of  the $e^{+}e^{-}\to\Sigma^{\pm}\bar{\Sigma}^{\mp}$ reaction~\cite{sigma}.  The free parameters in the fit are the normalization constants $c_{0}$ and $c_{1}$, which describes the average effect of a set of possible intermediate states representing the form factors in the framework of the vector-meson-dominance model~\cite{VMD}. The QCD scale $\Lambda_{\text{QCD}}$ is fixed to $0.3\gev$.  In the fit, both statistical and systematic uncertainties are taken into account.
The fit result is shown in Fig.~\ref{comgeff}~(a) with a fit quality of $\chi^{2}/{ndof}=9.98/5$, where $ndof$ is the number of degrees of freedom.
To obtain a better understanding of the full set of $\Sigma$ isospin states, the effective FFs of $\Sigma^{0}$ are compared with previous measurements of $\Sigma^{\pm}$ at BESIII~\cite{sigma} as shown in  Fig.~\ref{comgeff}~(b). An asymmetry in results is observed for the $\Sigma$ isospin triplet, with the $\Sigma^+$ results lying above the $\Sigma^0$ results which in turn are higher than the $\Sigma^-$ results.  This behavior confirms the hypothesis that
the effective FF is proportional to $\sum_{q} Q_{q}^{2}$ with $q=u,d,s$ quarks.
Moreover, the effective form factor of the $\Sigma^{0}$ is compared with that of the $\Lambda$ baryon to test the diquark correlation model. 
Our measurements are inconsistent with the hyperon-antihyperon ($Y\bar{Y}$) potential models from Ref.~\cite{int_fsi}, where the cross sections of $\ee\to\sigg$ exhibits a weaker energy dependence than the pQCD power-law function in Eq.~(\ref{pqcd}). 
We notice that there is a prediction for the non-resonant cross section of $\ee\to\sigg$ at the $J / \psi$ mass~\cite{jpsi}, based on an effective Lagrangian density, that is consistent with our result when extrapolated to $\sqrt{s}=3.097~\mathrm{GeV}$ using Eq.~(\ref{pqcd}).

\end{multicols}

\begin{multicols}{2}

\section{Summary}


In summary, data samples collected with the BESIII detector at c.m.~energies between $2.3864$~GeV and $3.0200$~GeV have been exploited to perform measurements  of the process $e^{+}e^{-}\to\Sigma^{0}\bar{\Sigma}^{0}$. 
Born cross sections and effective FFs have been determined with a novel method for c.m.~energies near production threshold at $\sqrt{s}=2.3864$ and $2.396$~GeV.
No significant signal is observed at $\sqrt{s}=$ $2.3864$$\gev$ 
and an upper limit on the cross section at the 90\% C.L. is determined. 
No threshold effect is observed for this process, in common with the behavior seen in an earlier analysis of $e^{+}e^{-}\to\Sigma^{\pm}\bar{\Sigma}^{\mp}$~\cite{sigma}.
In addition, a single-tag method has been applied at c.m.~energies between $2.3960$~GeV and $3.0200$~GeV. The measured results are in good agreement with previous results from BaBar~\cite{babar}, but the precision is significantly improved. The cross-section lineshape for $e^{+}e^{-}\to\Sigma^{0}\bar{\Sigma}^{0}$ is well described with a pQCD-motivated function. An asymmetry of the effective FFs of $\Sigma$ isospin triplet is observed, which is consistent with their incoherent sum of squared charges of valence quarks~\cite{sigma}. Our results also provide experimental inputs to test various theoretical models, such as $Y\bar{Y}$ potential~\cite{int_fsi} and diquark correlation~\cite{di} models.

\section*{Acknowledgements}
The BESIII collaboration thanks the staff of BEPCII and the IHEP computing center and the supercomputing center of USTC for their strong support. This work is supported in part by National Key Basic Research Program of China under Contract Nos. 2020YFA0406400, 2020YFA0406300; National Natural Science Foundation of China (NSFC) under Contracts Nos.~11625523, 11635010, 11735014, 11822506, 11835012, 11935015, 11935016, 11935018, 11961141012, 12035013, 11335008, 11375170, 11475164, 11475169, 11605196, 11605198, 11705192; the Chinese Academy of Sciences (CAS) Large-Scale Scientific Facility Program; Joint Large-Scale Scientific Facility Funds of the NSFC and CAS under Contracts Nos.~U1732263, U1832207, U1532102, U1732263, U1832103; CAS Key Research Program of Frontier Sciences under Contracts Nos.~QYZDJ-SSW-SLH003, QYZDJ-SSW-SLH040; 100 Talents Program of CAS; INPAC and Shanghai Key Laboratory for Particle Physics and Cosmology; ERC under Contract No.~758462; German Research Foundation DFG under Contracts Nos.~443159800, Collaborative Research Center CRC 1044, FOR 2359; Istituto Nazionale di Fisica Nucleare, Italy; Ministry of Development of Turkey under Contract No.~DPT2006K-120470; National Science and Technology fund; STFC (United Kingdom); The Knut and Alice Wallenberg Foundation (Sweden) under Contract No.~2016.0157; The Royal Society, UK under Contracts Nos.~DH140054, DH160214; The Swedish Research Council; Olle Engkvist Foundation under Contract No 200-0605; U. S. Department of Energy under Contracts Nos. DE-FG02-05ER41374, DE-SC-0012069.


\end{multicols}


\begin{thebibliography}{99}

\bibitem{hofstadter} R.~Hofstadter, Rev. Mod.~Phys. {\bf 28}, 214 (1956).

\bibitem{chiral} L.~S.~Geng, J.~Martin Camalich, L.~Alvarez-Ruso and M.~J.~Vicente Vacas, Phys.\ Rev.\ Lett.\  {\bf 101}, 222002 (2008).
\bibitem{pqcd}  S.~J.~Brodsky and G.~R.~Farrar, Phys.\ Rev.\ D {\bf 11}, 1309 (1975).
\bibitem{lattice1}  J.~R.~Green, J.~W.~Negele, A.~V.~Pochinsky, S.~N.~Syritsyn, M.~Engelhardt and S.~Krieg, Phys.\ Rev.\ D {\bf 90}, 074507 (2014). 

\bibitem{hyperon2} G.~Ramalho, K.~Tsushima and A.~W.~Thomas,  J.\ Phys.\ G {\bf 40}, 015102 (2013);
F.~Gross, G.~Ramalho and K.~Tsushima,  Phys.\ Lett.\ B {\bf 690}, 183 (2010).
\bibitem{hyperon1} G.~Eichmann, H.~Sanchis-Alepuz, R.~Williams, R.~Alkofer and C.~S.~Fischer, Prog.\ Part.\ Nucl.\ Phys.\  {\bf 91}, 1 (2016).

\bibitem{bes_fu} M.~Ablikim \textit{et al.} [BESIII Collaboration], Chin. Phys. C \textbf{44}, 040001 (2020).




\bibitem{time_babar} J.~P.~Lees {\it et al.} [BaBar Collaboration], Phys.\ Rev.\ D {\bf 87}, 092005 (2013).
\bibitem{time_cmd} R.~R.~Akhmetshin {\it et al.} [CMD-3 Collaboration], Phys.\ Lett.\ B {\bf 759}, 634 (2016).
\bibitem{lambdac_bes3} M.~Ablikim {\it et al.} [BESIII Collaboration], Phys.\ Rev.\ Lett.\ {\bf 120}, 132001 (2018).
\bibitem{neutron} M.~N.~Achasov {\it et al.} Phys.\ Rev.\ D {\bf 90}, 112007 (2014).
\bibitem{lamlam} M.~Ablikim {\it et al.} [BESIII Collaboration], Phys.\ Rev.\ D\ {\bf 97}, 032013 (2018); M.~Ablikim {\it et al.} [BESIII Collaboration], Phys. Rev. Lett. \textbf{123}, 122003 (2019).



\bibitem{int_fsi} J.~ Haidenbauer, U.-G.~Mei\ss{}ner,  and L.-Y.~Dai, Phys.\ Rev.\ D {\bf 103}, 014028 (2021).
\bibitem{int_res}  B.~El-Bennich, M.~Lacombe, B.~Loiseau and S.~Wycech, Phys.\ Rev.\ C {\bf 79}, 054001 (2009).


\bibitem{int_coulomb}  R.~Baldini~Ferroli, S.~Pacetti, A.~Zallo and A.~Zichichi, Eur.\ Phys.\ J.\ A {\bf 39}, 315 (2009);
R.~Baldini~Ferroli, S.~Pacetti and A.~Zallo, Eur.\ Phys.\ J.\ A {\bf 48}, 33 (2012).

\bibitem{sigma} M.~Ablikim {\it et al.} [BESIII Collaboration], Phys.\ Lett.\ B {\bf 814}, 136110 (2021).
\bibitem{cascas} M.~Ablikim {\it et al.} [BESIII Collaboration], Phys.\ Rev.\ D\ {\bf 103}, 012005 (2021); M.~Ablikim {\it et al.} [BESIII Collaboration], Phys.\ Lett.\ B {\bf 820}, 136557 (2021).
 
\bibitem{di} M. Anselmino {\it et al.}, Rev. Mod. Phys. {\bf 65}, 1199 (1993); R.~L.~Jaffe and F.~Wilczek, Phys.\ Rev.\ Lett.\  {\bf 91}, 232003 (2003);  R.~L.~Jaffe, Phys.\ Rept.\  {\bf 409}, 1 (2005).

\bibitem{babar} B.~Aubert {\it et al.} [BaBar Collaboration], Phys.\ Rev.\ D\ {\bf 76}, 092006 (2007).

\bibitem{dataset} M.~Ablikim {\it et al.} [BESIII Collaboration], Chin.\ Phy.\ C {\bf 41}, 063001 (2017).

\bibitem{Ablikim:2009aa} M.~Ablikim {\it et al.} [BESIII Collaboration], Nucl.\ Instrum.\ Meth.\ A {\bf 614}, 345 (2010).
Methods {\bf 1}, 15 (2017)


\bibitem{geant4} S.~Agostinelli {\it et al.} [GEANT4 Collaboration], Nucl.\ Instrum.\ Meth.\ A {\bf 506}, 250 (2003).

\bibitem{conexc} R. G. Ping, Chin.\ Phys.\ C {\bf 38}, 083001 (2014).

\bibitem{Anjrej} P.~Elisabetta, F.~G\"oran, K.~Andrzej, L.~Stefan and S.~Jiao Jiao, Phys. Rev. D {\bf 99}, 056008 (2019); F.~G\"oran and S.~Karin, Phys. Rev. D {\bf 101}, 033002 (2020).

\bibitem{bes} D.~J.~Lange, Nucl.\ Instrum\ Meth.\ A {\bf 462}, 152 (2001); R.~G.~Ping {\it et al.} Chin.\ Phys. C {\bf 32}, 599 (2008).

\bibitem{pdg} P.~A.~Zyla {\it et al.} [Particle Data Group], Prog.\ Theor.\ Exp. Phys. {\bf 2020}, 083C01 (2020).


\bibitem{lundarlw} B.~Andersson and H.~Hu, hep-ph/9910285.
\bibitem{babayaga} G.~Balossini {\it et al.}, Nucl.\ Phys.\ B {\bf 758}, 227 (2006);
G.~Balossini {\it et al.}, Phys.\ Lett.\ B {\bf 663}, 209 (2008).

\bibitem{bestwogam}  S.~Nova, A.~Olchevski and T.~Todorov, DELPHI-90-35 PROG 152 (1990).
S. Nova, A.~Olchevski and T.~Todorov [DELPHI Collaboration], DELPHI 90-35 PROG {\bf 152} 1990.

\bibitem{ee} N.~Cabibbo and R.~Gatto, Phys.\ Rev.\ Lett.\ {\bf 4}, 313 (1960);\ Phys.\ Rev.\ {\bf 124}, 1577 (1961).

\bibitem{schwinger} J.~Schwinger, $Particle,~Sources,~and~Field$~(Perseus Books Publishing, Massachusetts, 1998), Vol.~3.
\bibitem{coulomb}  A.~B.~Arbuzov and T.~V.~Kopylova, JHEP {\bf 1204}, 009 (2012).


\bibitem{hyperon_babar} B.~Aubert {\it et al.} [BaBar Collaboration], Phys.\ Rev.\ D {\bf 76}, 092006 (2007).


\bibitem{VP} S.~Actis {\it et al.} [Working Group on Radiative Corrections and Monte Carlo Generators for Low Energies], Eur.\ Phys.\ J.\ C {\bf 66}, 585 (2010).

\bibitem{ana_tpo} X.~Y.~Zhou, S.~X.~Du, G.~Li and C.~P.~Shen, Comput. Phys. Commun. {\bf 258}, 107540 (2021).





\bibitem{bays} K. Stenson, arXiv:physics/0605236; J. Lundberg, J. Conrad, W. Rolke, and A. Lopez, Comput. Phys. Commun. {\bf 181}, 683 (2010).

\bibitem{wilk} S.S. Wilks, {\it The large-sample distribution of the likelihood ration for testing composite hypotheses}, Ann. Math. Stat., 9, (1938), pp. 60-62.

\bibitem{sys} R. Wanke, {\it Data Analysis in High Energy Physics}, (Wiley-VCH Verlag GmbH and Co. KGaA, Singapore, 2013), pp. 263$-$280.

\bibitem{vertex} M.~Xu {\it et al.}, Chin.\ Phys.\ C {\bf 33}, 428 (2009).



\bibitem{photon} M.~Ablikim {\it et al.} [BESIII Collaboration], Phys.\ Rev.\ D {\bf 92},  052003 (2015).






\bibitem{rinaldo} S.~Pacetti, R.~Baldini~Ferroli and E.~Tomasi Gustafsson, Phys. Rept. {\bf 550-551}, 1 (2015).
\bibitem{VMD} H.~W.~Hammer, U.~G.~Meissner and D.~Drechsel, Phys.\ Lett.\ B {\bf 385}, 343 (1996).

\bibitem{jpsi} R.~Baldini Ferroli, A.~Mangoni, S.~Pacetti and K.~Zhu, Phys.\ Lett.\ B {\bf 799}, 135041 (2019).


\end{thebibliography}
\end{document}